\documentclass[11pt,a4paper]{article} 
\pdfoutput=1
\usepackage{jheppub,hyperref,mdwlist}
\usepackage{amsmath}
\usepackage{graphicx}
\usepackage{subfigure}
\usepackage{amssymb}
\usepackage{xcolor}
\usepackage{multirow}
\usepackage{cancel}
\usepackage{color}
\usepackage{listings}
\usepackage{verbatim}
\usepackage{float}
\usepackage{slashed}
\usepackage{mathtools}
\usepackage{soul}

\usepackage{tikz}
\usetikzlibrary{arrows,shapes}
\usetikzlibrary{trees}
\usetikzlibrary{matrix,arrows} 				
\usetikzlibrary{positioning}				
\usetikzlibrary{calc,through}				
\usetikzlibrary{decorations.pathreplacing}  
\usepackage{pgffor}							

\usetikzlibrary{decorations.pathmorphing}	
\usetikzlibrary{decorations.markings}
\tikzset{
    vector/.style={decorate, decoration={snake}, draw},
	provector/.style={decorate, decoration={snake,amplitude=2.5pt}, draw},
	antivector/.style={decorate, decoration={snake,amplitude=-2.5pt}, draw},
    fermion/.style={draw=black, postaction={decorate},
        decoration={markings,mark=at position .55 with {\arrow[draw=black]{>}}}},
    fermionbar/.style={draw=black, postaction={decorate},
        decoration={markings,mark=at position .55 with {\arrow[draw=black]{<}}}},
    fermionnoarrow/.style={draw=black},
    gluon/.style={decorate, draw=black,
        decoration={coil,amplitude=4pt, segment length=5pt}},
    scalar/.style={dashed,draw=black, postaction={decorate},
        decoration={markings,mark=at position .55 with {\arrow[draw=black]{>}}}},
    scalarbar/.style={dashed,draw=black, postaction={decorate},
        decoration={markings,mark=at position .55 with {\arrow[draw=black]{<}}}},
    scalarnoarrow/.style={dashed,draw=black},
    electron/.style={draw=black, postaction={decorate},
        decoration={markings,mark=at position .55 with {\arrow[draw=black]{>}}}},
	bigvector/.style={decorate, decoration={snake,amplitude=4pt}, draw},
}

\tikzstyle{block} = [draw, rectangle, 
    minimum height=3em, minimum width=6em]

\usepackage{amsmath}
\usepackage{wrapfig}
\usepackage{cleveref}

\newcommand{\be}{\begin{equation}}
\newcommand{\ee}{\end{equation}}
\newcommand{\beq}{\begin{equation}}
\newcommand{\eeq}{\end{equation}}
\newcommand{\bea}{\begin{eqnarray}}
\newcommand{\eea}{\end{eqnarray}}
\newcommand{\besp}{\begin{equation}\begin{split}}
\newcommand{\eesp}{\end{split}\end{equation}}

\newcommand{\nn}{\nonumber}

\newcommand{\Eq}[1]{Eq.~(\ref{#1})}

\newcommand{\Dfbd}{\mathord{\buildrel{\lower3pt\hbox{$\scriptscriptstyle\leftrightarrow$}}\over {D}_{\mu}}}

\def\mC{\mathcal{C}}
\def\mD{\mathcal{D}}

\def\mF{\mathcal{F}}

\def\mI{\mathcal{I}}

\def\mL{\mathcal{L}}
\def\mM{\mathcal{M}}

\def\mO{\mathcal{O}}
\def\mP{\mathcal{P}}

\def\mR{\mathcal{R}}

\def\vec{\mathbf}

\def\0{\textbf{0}}
\def\1{\textbf{1}}
\def\2{\textbf{2}}
\def\3{\textbf{3}}
\def\4{\textbf{4}}
\def\5{\textbf{5}}
\def\6{\textbf{6}}
\def\7{\textbf{7}}
\def\8{\textbf{8}}
\def\9{\textbf{9}}

\def\d{\text{d}}

\def\p{\textbf{p}}

\definecolor{RoyalBlue}{cmyk}{1, 0.50, 0, 0}

\begin{document}

\title{Photon proliferation from multi-body dark matter annihilation}

\author[a]{Shao-Ping Li}
\affiliation[a]{Department of Physics, The University of  Osaka, Toyonaka, Osaka 560-0043, Japan}

\author[b]{and Ke-Pan Xie}

\affiliation[b]{School of Physics, Beihang University, Beijing 100191, China}

\emailAdd{lisp@het.phys.sci.osaka-u.ac.jp}
\emailAdd{kpxie@buaa.edu.cn}

\abstract{
Multi-body dark matter  annihilation is commonly expected to be suppressed by higher-order couplings and phase-space factors, therefore being ignored thus far. We show that, however, this does not hold for a class of nonthermal dark matter scenarios, where the dark matter particle becomes nonrelativistic at temperatures much higher than its mass.  We exemplify such a multi-body process via ultralight pseudoscalar dark matter annihilation to diphotons, which leads to a novel photon proliferation effect in the early Universe. As a phenomenological application, we consider the photon temperature shift after neutrino decoupling, showing that the photon proliferation effect can render bounds on the ultralight dark matter couplings stronger than the existing constraints by several orders of magnitude. Our research can be extended to other interactions and dark matter candidates,  highlighting the importance of multi-body processes in the early Universe.
}

\maketitle
\flushbottom

\section{Introduction}

Dark matter (DM) contributes approximately $84\%$ of the total matter in the Universe~\cite{Planck:2018vyg}, yet its particle nature remains mysterious.  Numerous particle DM models have emerged in recent decades. These models typically involve interactions between DM and the Standard Model (SM), allowing DM annihilation into SM particles and opening detectable channels. In general, the $2\to2$ processes, in which DM annihilates into $e^+e^-$, $\gamma\gamma$, or $\nu\bar\nu$, have become key multi-messenger probes in modern cosmology and astrophysics, with potential signals arising from the diffuse photon background~\cite{Bringmann:2012ez,Hooper:2012sr}, cosmic rays~\cite{Hisano:2009rc,Fermi-LAT:2009ppq}, cosmic microwave background (CMB) anisotropies~\cite{Padmanabhan:2005es,Chluba:2009uv}, CMB spectral distortions~\cite{Chluba:2011hw,Chluba:2016bvg,Li:2024xlr}, big-bang nucleosynthesis (BBN)~\cite{Jedamzik:2009uy,Kawasaki:2015yya,Depta:2019lbe}, and extra neutrino abundances~\cite{Boehm:2013jpa,Nollett:2014lwa,Arguelles:2019ouk}.

While the $3\to 2$ and $4\to 2$ processes may also play an important role in certain contexts~\cite{Carlson:1992fn,Hochberg:2014dra,Bernal:2015xba},
$N$-body DM annihilation $N\to 2$ with $N\geqslant 5$ is rarely considered. Such a dearth of investigations  comes from the common expectation  that   the squared amplitudes of  $N\to 2$ annihilation would be  suppressed by  higher-order weak couplings or heavy mediator particles, and the final annihilation rate would also be suppressed by multi-body phase-space factors.  Nevertheless, the physical impacts of DM annihilation on astrophysics and   cosmology depend not only on the squared amplitude, but also on  DM kinematics  and the background DM  number density.

For an $N\to 2$ DM annihilation process, for example, the particle number changing rate of the annihilation products scales as
\begin{align}
\frac{\d n}{\d t}\propto \int \left(\prod_{i=1}^N\frac{\d^3p_i}{(2\pi)^32E_i}f_i\right)|\mM_N|^2\,,
\end{align}
where $f_i=f_{\rm dm}(\p_i)$ denotes the DM distribution function with $\p_i$ the momentum of the $i$-th DM particle. The form of the squared amplitude $|\mM_N|^2$ depends  on specific particle physics interactions, which is typically suppressed by higher-order weak couplings. If the $N\to 2$ process occurs at a temperature $T$ when DM is already {\it nonrelativistic}, we may parametrize $|\mathcal{M}_N|^2\propto g^{2N}/m_{\rm dm}^{2N-4}$, where $g$ denotes some dimensionless DM coupling and the energy scale is determined by the DM mass. In this case, the annihilation rate scales as  $\d n/\d t\sim [ g^{2}(T/m_{\rm dm})^{3}(\text{eV}/m_{\rm dm})]^N m_{\rm dm}^4$, where the $N$-body phase-space integration on $f_i$ yields the DM number density $n_{\rm dm}\sim (T/\text{eV})^3(\text{eV}/m_{\rm dm})~{\rm eV}^3$. Bearing the suppression factor $g^{2}$, the annihilation rate for a given DM mass may still be boosted by a large $N$ at sufficiently  high temperatures $T\gg m_{\rm dm}$.   

For DM production from thermal particles, where DM typically has the  temperature and momentum dependence similar to  thermal particles,  DM  such as the weakly interacting massive particles (WIMPs) typically becomes nonrelativistic at $T\lesssim m_{\rm dm}$. In this case, the enhancement in the  high-temperature regime is limited. However,  nonthermal DM production can feature much colder DM that becomes nonrelativistic at $T\gg m_{\rm dm}$, such that  the annihilation rate would receive larger enhancements in a higher temperature regime. Known examples of such nonthermal DM production include misalignment mechanism~\cite{Abbott:1982af,Dine:1982ah,Preskill:1982cy} for ultralight   DM~\cite{Arias:2012az,DiLuzio:2020wdo,Ferreira:2020fam,Graham:2015ouw}, axion or vector boson DM production from parametric resonance~\cite{Co:2017mop,Dror:2018pdh} and inflationary
fluctuations~\cite{Graham:2015rva}, as well as superweakly or feebly interacting massive DM from nonthermal particle decay or annihilation~\cite{Feng:2003xh,Feng:2003uy,Hall:2009bx}.

\begin{figure}[t]
	\centering
	\includegraphics[scale=0.55]{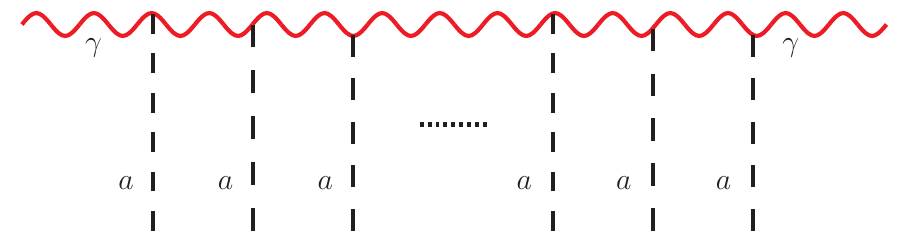} 
	\caption{\label{fig:NDM2}$N$-body DM annihilation to diphotons via the $a$-$\gamma$-$\gamma$ vertex, where $a$ denotes a pseudoscalar DM particle.}
\end{figure}

The above discussions show that multi-body DM annihilation for a broad class of nonthermal DM scenarios  can  be more important than the conventional $2\to 2$ process in the early Universe. Depending on the annihilation product, the boosted multi-body DM annihilation rate can cause significant impacts on   cosmological observables inherited from much earlier times, such as the initial density perturbations, light elements in BBN,  primordial CMB spectral distortions,  extra  energy components,  and the  frozen DM relic density. The enhanced impacts can on the one hand render bounds on DM interactions more severe,  and on the other hand open new channels for  probing hidden particles in the early Universe. 

In this work, we exemplify such a multi-body process via ultralight pseudoscalar DM annihilation to diphotons, where a novel  photon proliferation effect arises. The annihilation channel is present in a broad class of ultralight DM scenarios, typically via the \textit{fence} diagram shown in Fig.~\ref{fig:NDM2}. The injected photons can lead to important physical consequences at redshifts much higher than the recombination epoch ($z\gtrsim1100$), including modification to the effective neutrino number~\cite{Escudero:2018mvt,Bennett:2019ewm,EscuderoAbenza:2020cmq,Drewes:2024wbw}, discrepancy of the baryon asymmetry between the BBN and CMB~\cite{Yeh:2022heq}, and the primordial CMB spectral distortions~\cite{Hu:1992dc,Chluba:2013kua}. The illustration presented in this work can also be extended beyond the diphoton channel and applied to other interactions, such as the DM-neutrino interactions, as well as other DM candidates, such as the dark photon~\cite{Caputo:2021eaa,Fabbrichesi:2020wbt}. It highlights the importance of general multi-body annihilation processes in the early Universe.

\section{Photon proliferation}

\subsection{Photon energy injection}\label{sec:gamma-einj}

We will illustrate the photon proliferation effect via a light pseudoscalar DM particle with mass below the eV scale, where the $N\to 2$ process dates back to neutrino decoupling in the early Universe. While the photon-number injection rate  is larger than the energy release rate in this  regime~\cite{Li:2024xlr}, the net effect from photon-number injection still acts as the energy release to shift the background photon temperature~\cite{Chluba:2013kua}  after neutrino decouples at $T=\mathcal{O}(1)$~MeV, or to cause  primordial spectral distortions after $T\lesssim 1$~keV~\cite{Chluba:2015hma}. For ultralight DM, the number of particles (occupancy) within the volume of the Compton wavelength,  $1/m^3_{\rm dm}$, is typically large in the early Universe, $n_{\rm dm}/m_{\rm dm}^3\gg 1$,  such that the set of DM particles is usually described by classical field~\cite{Hu:2000ke,Hui:2016ltb,Hui:2021tkt}. Here, we will work on the particle collision pattern to describe DM interactions. We expect that both approaches should give the same result after proper matching, as already hinted by the equivalence between the Boltzmann equation and classical field theory~\cite{Mueller:2002gd,Jeon:2004dh,Hertzberg:2016tal,Li:2025wwl}.

The evolution of photon energy is determined by the general Boltzmann equation, $\d\rho_\gamma/\d t+4H\rho_\gamma=\mathcal{C}$, with $\mathcal{C}$ the energy collision rate.  Defining $\xi_\gamma\equiv (\rho_\gamma-\rho_{\rm bg})/\rho_{\rm bg}$, we can write down the evolution of $\xi_\gamma$ as
\begin{align}
\frac{\d\xi_\gamma}{\d t}=\frac{1}{\rho_{\rm bg}}\sum_{N=2}^{\infty}\mathcal{C}_N\,,
\end{align} 
where we have used the relation $\d\rho_{\rm bg}/\d t+4H\rho_{\rm bg}=0$ for the background photon energy density $\rho_{\rm bg}\approx 0.66\,T^4$, and $\mathcal{C}_N$ denotes  the energy collision rate from DM $N\to 2$ annihilation and photon $2\to N$ coalescence
\be\label{difference}
\mC_N=2\int\d\Phi_{\rm dm}\d\Phi_\gamma (2\pi)^4\delta^{4}(p)\frac{|\tilde\mM_N|^2}{2N!}E_a\mathcal{F}\,,
\ee
where the phase-space volume elements are 
\begin{align}
\d\Phi_{\rm dm}\equiv \prod_{i=1}^N\frac{\d^3p_i}{(2\pi)^32E_i}\,,\quad \d\Phi_\gamma\equiv  \frac{\d^3p_a}{(2\pi)^32E_a}\frac{\d^3p_b}{(2\pi)^32E_b}\,,
\end{align} 
and $\delta^4(p)\equiv \delta^4 (p_a+p_b-\sum_{i=1}^Np_i)$. 
The distribution function $\mathcal{F}$ reads
\be\label{eq:F-def}
\mathcal{F}\equiv (1+f'_a)(1+f'_b)\left(\prod_{i=1}^Nf_i\right)-f_af_b\prod_{i=1}^N(1+f'_i)\,,
\ee
where $f_i^{(\prime)}\equiv f_{\rm dm}^{(\prime)}(\p_i)$ and $f_{a,b}^{(\prime)}\equiv f_\gamma^{(\prime)}(\p_{a,b})$ with $f_{\rm dm}^{(\prime)}$ and $f_\gamma^{(\prime)}$ are the DM and photon phase-space distribution functions, respectively. The symmetry factor $1/(2N!)$ is shown explicitly due to the identical photons and DM particles.

The photons injected from $N\to 2$ are nonthermal, though they will quickly be thermalized with the background photons; while photons in the $2\to N$ process come from the thermal plasma. The initial-state DM particles are nonthermal, nonrelativistic, and abundant, while the produced DM particles are rare and relativistic. Therefore, we can set $f'_{\rm dm}=0$ and $f'_\gamma=0$, while $f_{\rm dm}(\p)\approx 0$ when $|\p|\gtrsim m_{\rm dm}$, and $f_\gamma(\p)=1/(e^{|\p|/T}-1)$. Consequently,  \Eq{eq:F-def} reduce to
\be
\mathcal{F}\approx \left(\prod_{i=1}^Nf_i\right)-f_af_b\,.
\ee
The squared amplitude $|\tilde\mM_N|^2$ is model dependent. For general discussion, we will remain agnostic on the production mechanism for ultralight DM~\cite{Abbott:1982af,Dine:1982ah,Preskill:1982cy,Co:2017mop,Dror:2018pdh,Graham:2015rva,Feng:2003xh,Feng:2003uy,Hall:2009bx}; we  assume $|\tilde\mM_N|^2$ can be parameterized as $\approx\alpha_N s^{\beta_N}$, where $\sqrt{s}$ is the energy in the center-of-mass frame, and $(\alpha_N,\beta_N)$ are determined by the specific annihilation process. For example, if DM only couples to photons via $-g_{a\gamma\gamma}aF_{\mu\nu}\tilde F^{\mu\nu}/4$, then the dimensionful $\alpha_N$ scales as $\alpha_N\propto g_{a\gamma\gamma}^{2N}$ while the power reads $\beta_N=2$. If the DM self-interaction $-\lambda a^4/4!$ exists and dominates the annihilation, then $\alpha_N\propto \lambda^{N-1}g_{a\gamma\gamma}^2$ and $\beta_N=3-N$. We will only consider the case of $\beta_N\leqslant2$. Furthermore, we consider the following scales
\be\label{eq:conditions}
T\sim1~{\rm MeV},\quad m_{\rm dm}\ll 1~{\rm eV},\quad Nm_{\rm dm}\ll T.
\ee

\Eq{difference} can be treated as the difference between two positive-definite terms, $\mC_N\equiv\mD_N-\mP_N$, according to the expressions in $\mF$. We first calculate
\be
\mD_N\approx2\int\d\Phi_{\rm dm}\d\Phi_\gamma (2\pi)^4\delta^{4}(p)\frac{|\tilde\mM_N|^2}{2N!}E_a\left(\prod_{i=1}^Nf_i\right)\,.
\ee
A remarkable feature of this integration is that each $\p_i$ is dressed with a nonrelativistic distribution $f_i$. Therefore, we can make the replacement $p_i^\mu\to(m_{\rm dm},0,0,0)$, and hence $\delta^4(p_a+p_b-\sum_{i=1}^Np_i)\to\delta(E_a+E_b-Nm_{\rm dm})\delta^{(3)}(\p_a+\p_b)$ becomes irrelevant to $p_i$. The integration on $p_i$ can then  be performed trivially in the nonrelativistic regime, with  $E_a\to N m_{\rm dm}/2$, 
\be\label{eq:NR}
\int\frac{\d^3p_i}{(2\pi)^3}f_i\to n_{\rm dm},\quad |\tilde\mM_{N}|^2\xrightarrow[\sqrt{s}\to Nm_{\rm dm}]{\text{nonrelativistic}}\alpha N^{2\beta}m_{\rm dm}^{2\beta}\equiv|\mM_{N}|^2,
\ee
and hence
\be
\mD_N\approx\frac{N m_{\rm dm}}{16\pi N!}\left(\frac{n_{\rm dm}}{2m_{\rm dm}}\right)^N|\mM_N|^2\,,
\ee
which is the  contribution from $N\to2$ DM annihilation.

Next we calculate the contribution from the $2\to N$ photon coalescence, i.e.,
\be
\mP_N\approx2\int\d\Phi_{\rm dm}\d\Phi_\gamma (2\pi)^4\delta^{4}(p)\frac{|\tilde\mM_N|^2}{2N!}E_af_af_b\,.
\ee
As not all $p_i$ are associated with a $f_i$,  we cannot  make the replacement of \Eq{eq:NR}. In particular, some of DM  momenta can be in the relativistic regime such that $s\gg N^2 m_{\rm dm}^2$. 
Consequently, we have to use $|\tilde\mM_N|^2$ instead of $|\mM_N|^2$, and the calculation of $\mP_N$ will be rather involved. However, an upper limit of $\mP_N$ can still be derived, and we are able to show that $\mP_N\ll \mD_N$ for $N\geqslant4$. 

Let $\textbf{P}=\sum_{i=1}^N\p_i$ and $E=\sum_{i=1}^NE_i$ (such that $s=E^2-P^2$). We can first complete the integration on $p_{a,b}$ and obtain
\be
\mP_N=
\int\left(\prod_{i=1}^N\frac{\d^3p_i}{(2\pi)^32E_i}\right)\frac{|\tilde\mM_{N}|^2}{2N!}\mI,
\ee
where 
\be\label{eq:IP}\begin{split}
\mI=&~2\int\frac{\d^3p_a}{(2\pi)^32E_a}\frac{\d^3p_b}{(2\pi)^32E_b}(2\pi)^4\delta^3(\p_a+\p_b-\textbf{P})\delta(E_a+E_b-E)E_af_af_b\\[0.2cm]
=&~\frac{E}{8\pi\left(e^{E/T}-1\right)}\left[\frac{2T}{P}\log\left(\frac{e^{\frac{E+P}{2T}}-1}{e^{\frac{E-P}{2T}}-1}\right)-1\right],
\end{split}\ee
is a function of $E$ and $P=|\textbf{P}|$.
Define the ratio
\begin{multline}
\mR_N=\frac{\mP_N}{\mD_N}\approx\frac{1}{Nm_{\rm dm}}\left(\frac{2m_{\rm dm}}{n_{\rm dm}}\right)^N\int\left(\prod_i\frac{\d^3p_i}{(2\pi)^32E_i}\right)\\[0.2cm]
\times\frac{E}{e^{E/T}-1}\left[\frac{2T}{P}\log\left(\frac{e^{\frac{E+P}{2T}}-1}{e^{\frac{E-P}{2T}}-1}\right)-1\right]\left(\frac{E^2-P^2}{N^2m_{\rm dm}^2}\right)^{\beta_N},
\end{multline}
and attempt to set an upper limit for it. Notice that $\sqrt{E^2-P^2}>Nm_{\rm dm}$, and hence the base of the power $(\cdots)^{\beta_N}$ is larger than 1. As a result, $\mR_N\leqslant \mR_N|_{\beta_N=2}$. Another important fact is that when fixing $E$, the following function
\be
F(P)=\left[\frac{2T}{P}\log\left(\frac{e^{\frac{E+P}{2T}}-1}{e^{\frac{E-P}{2T}}-1}\right)-1\right](E^2-P^2)^2,
\ee
is monotonically decreasing with $P$, and the boundary conditions are $F(E)=0$, $F(0)=E^4\coth(E/4T)$. Therefore, we reach an upper limit of
\be
\mR_N\leqslant\left(\frac{2m_{\rm dm}}{n_{\rm dm}}\right)^N\left(\frac{1}{Nm_{\rm dm}}\right)^5\int\left(\prod_{i=1}\frac{\d^3p_i}{(2\pi)^32E_i}\right)\frac{E^5}{e^{E/T}-1}\coth\left(\frac{E}{4T}\right).
\ee
Although the integration on $p_i$ is not easy to evaluate, the factor $\sim E^5\coth(E/4T)/(e^{E/T}-1)$ is maximized at $E\sim 9T/2$ with a value around $8\pi T^5$. Therefore, as a conservative estimate, we have 
\be\label{R-final}\begin{split}
\mR_N\lesssim&~8\pi\left(\frac{2m_{\rm dm}}{n_{\rm dm}}\right)^N\left(\frac{T}{Nm_{\rm dm}}\right)^5\int_{E<9T/2}\left(\prod_{i=1}\frac{\d^3p_i}{(2\pi)^32E_i}\right)\\
\lesssim&\left(\frac{2m_{\rm dm}}{n_{\rm dm}}\right)^N\left(\frac{T}{Nm_{\rm dm}}\right)^5\frac{8\pi}{(2N)!}\left(\frac{9T}{4\pi}\right)^{2N}.
\end{split}\ee
Given \Eq{eq:conditions}, we find $\mR_N\ll1$ is always satisfied for $N\geqslant4$, and hence the corresponding $\mP_N$'s can be dropped when calculating $\mC_N$.

Summarizing the available results, we get $\d\xi_\gamma/\d t=(\sum_{N=2}^\infty\mD_N-\mP_2-\mP_3)/\rho_{\rm bg}$. As will be shown later, in the processes we consider, $N=\mO(100)$ dominates the summation of $\mD_N$, where $\mD_N\gg\mP_2$, $\mP_3$ always holds. Therefore, the $\mP_{2,3}$ terms can be further omitted, and eventually we have reduced \Eq{difference} to
\begin{align}\label{eq:net_deltan/deltat}
	\frac{\d\xi_\gamma}{\d t}\approx\sum_{N=2}^\infty\frac{N m_{\rm dm}|\mM_{N}|^2}{16\pi N! \rho_{\rm bg}}\left(\frac{n_{\rm dm}}{2m_{\rm dm}}\right)^N.
\end{align}
We anticipate from this equation that the enhancement   from  $(n_{\rm dm}/m_{\rm dm})^N$  may dominate over the suppression from higher-order couplings in $|\mathcal{M}_N|^2$ up to certain $N$,  leading to an enhancement of $\xi_\gamma$.
When $N$ further increases, however, the $1/N!$  factor becomes important to suppress $\xi_\gamma$.  Consequently, we reach a   conclusion that   the dominant contribution comes from  $N\to 2$ annihilation instead of the conventional $2\to2$ channel, and a significant amount of photon energy will be generated. 

\subsection{The modification to the effective neutrino number }

Integrating Eq.~\eqref{eq:net_deltan/deltat} yields the total  photon energy release
\be\label{deltan}
\delta\xi_\gamma=\int_{z_1}^{z_2}\d z\left(\frac{\d\xi_\gamma}{\d t}\right)\left|\frac{\d t}{\d z}\right|\,,
\ee
where  $z_2\approx5.6\times10^9$ corresponds to the neutrino decoupling temperature $T\approx1.32$ MeV~\cite{Drewes:2024wbw}, and we have cut the lower integration limit at $z_1\approx2\times10^6$, corresponding  to the dawn of the primordial CMB $\mu$-distortion formation~\cite{Hu:1992dc,Chluba:2013kua}. Extending the integration limit to lower redshifts   causes negligible changes since the photon-energy release rate is high-temperature dominated. 

In the presence of a photon temperature shift, the effective neutrino number, 
\begin{align}
N_{\rm eff}\equiv \frac{8}{7}\left(\frac{11}{4}\right)^{4/3}\frac{\rho_\nu}{\rho_\gamma}\,,
\end{align}
with $\rho_{\nu}, \rho_\gamma$ the neutrino and photon energy densities, is modified by $\Delta N_{\rm eff}=N_{\rm eff}-N_{\rm eff}^{\rm SM}\approx-N_{\rm eff}^{\rm SM}\delta\xi_\gamma$ with $N_{\rm eff}^{\rm SM}$ the SM prediction. The resulting $\Delta N_{\rm eff}$ can be analytically written as
\begin{align}\label{eq:DN_numerical}
\Delta N_{\rm eff}\approx\sum_{N=2}^\infty c_N
\left(\frac{|\mM_{N}|^2}{m_{\rm dm}^{4-2N}}\right) \left(\frac{3.6\times10^4~{\rm eV}}{m_{\rm dm}}\right)^{4N-5}\,,
\end{align}
with $c_N\equiv -1.4\times10^{12}/[(N-1)!(N-2)]$. It is worthwhile to emphasize that under the infinite summation over $N$, the final result is convergent, as will be specified via some examples shown in section~\ref{sec:bounds}.

The above modification must be subject to the joint bound from the  BBN and CMB measurements: $|\Delta N_{\rm eff}|<0.429$ at $2\sigma$ level~\cite{Yeh:2022heq}. For numerical discussions, we will take the instantaneous neutrino decoupling limit with $N_{\rm eff}^{\rm SM}=3$. While going beyond the instantaneous decoupling leads to a different prediction of $N_{\rm eff}^{\rm SM}$ at $4\%$ level~\cite{Escudero:2018mvt,Bennett:2019ewm,EscuderoAbenza:2020cmq,Drewes:2024wbw}, the constraint presented below would not change noticeably. Additionally, using non-universal decoupling temperatures for three neutrinos would only cause a slight change of the bound. 
While we have used a neutrino decoupling temperature around 1~MeV, the large photon proliferation effect above 1~MeV   also  leads to modification of the Hubble expansion and the  followed neutrino decoupling, where constraints are expected to be somewhat stronger. 
These refined bounds can be obtained straightforwardly by substituting \Eq{eq:net_deltan/deltat} into the full neutrino Boltzmann equation, though it will not be pursued here.
As a phenomenological study, we show in  the next section  that  for ultralight pseudoscalar DM~\cite{Arias:2012az,DiLuzio:2020wdo,Ferreira:2020fam,Graham:2015ouw} the photon proliferation effect significantly strengthens current constraints on     DM couplings by several orders of magnitude, thereby excluding a substantial portion of the parameter space targeted by future experiments. 

\Eq{deltan} does not include the final-state photon Bose-Einstein enhancement in calculating the produced photons from  ultralight DM $N\to2$ annihilation. Such an effect has a close relation to parametric resonance (PR) production of photons. We comment here on the difference between photon production from $N\to 2$ annihilation and from PR.  Ultralight DM can also be treated as an oscillating classical field, which can yield PR as a non-perturbative particle production in a time-varying background~\cite{Kofman:1994rk,Kofman:1997yn}. In recent years, PR has also been described by  the method of Boltzmann equations based on the particle decay perspective~\cite{Alonso-Alvarez:2019ssa,Carenza:2019vzg,Moroi:2020bkq,Li:2025wwl}. For narrow PR, it was shown that Boltzmann equation can give quantitatively consistent prediction~\cite{Li:2025wwl}, if the Hubble expansion and background interactions can be neglected.  This can be interpreted as the  stimulated decay caused by final-state Bose-Einstein enhancement, which leads to exponential growth of particle occupation number. For broad PR~\cite{Fujisaki:1995dy,Fujisaki:1995ua,Khlebnikov:1996wr,Dufaux:2006ee}, similar exponential growth based on semi-analytical approaches also appears~\cite{Yoshimura:1995gc,Dufaux:2006ee},\footnote{Note that, however, these semi-analytical results still do not reach consistency among themselves, and can only be applied under certain approximations.} though it is currently not clear yet about the underlying quantum physics responsible for the abundant particle production, and is still an ongoing topic on the theoretical front.  In particular, unlike the narrow case, whether broad PR can be understood as  stimulated decay remains to be seen.  

For both narrow and broad PRs, whether the particle occupation number develops exponential growth depends sensitively on the produced momentum band. In the narrow regime, the resonantly produced photons populate at $|{\bf k}|=m_{\rm dm}/2$, and the bandwidth is given by $\delta |{\bf k}|\simeq q m_{\rm dm}\ll m_{\rm dm}/2$, where $q\ll1$ denotes the narrow regime. A small momentum bandwidth is fragile when exposed to  the redshift effect, and more importantly to the interactions between produced photons and the thermal plasma~\cite{Kasuya:1996np,Kofman:1997yn,Alonso-Alvarez:2019ssa,Carenza:2019vzg,Jaeckel:2021xyo}. This makes it different from preheating process right after inflation, where  no thermal plasma has established and hence PR can be  significant. In the broad regime, the momentum bandwidth scales approximately as $\delta |{\bf k}|\simeq  \sqrt{q}e^{-\sqrt{q}}m_{\rm dm}$~\cite{Dufaux:2006ee}, and again, suffers from any effect that drags  or redistributes $|{\bf k}|\sim m_{\rm dm}$ away from the resonant band. Once these momentum changes occur, the exponential growth will be terminated or  never be developed. 
At cosmic temperatures well before the photon last scattering, the background photons have tight contacts with hot electrons and nucleons, where the photons have a mean free path much smaller than the Hubble length.  It implies that  the produced photons will  basically undergo \textit{instantaneous} redistribution with the background plasma, such that plasma interactions dominate over the Hubble redshift effects to redistribute the produced photon momentum, rendering exponential growth of PR ineffective. Different from PR production that requires stable momentum bands, photons produced during $N\to2$ processes will directly have impact on the background plasma, even if the  redistribution with the background plasma is instantaneous. In particular, the produced soft photons $|{\bf k}|\sim m_{\rm dm}\ll T$ at $T\gtrsim 1$~keV  will be quickly absorbed by, e.g., double Compton scattering with the background electrons ($e_{\rm bg}$): $\gamma+e_{\rm bg}\to e+ \gamma+\gamma$,\footnote{Double Compton scattering has a rate much larger than the Hubble expansion rate by orders of magnitude at $T>1$~keV.} causing a temperature shift on the background photons. Based on the above considerations, we will not  include the PR effects in our numerical calculation.

\section{Bounds on ultralight DM couplings}\label{sec:bounds}

\subsection{The effective DM-photon interaction}\label{sec:gagg}

Let us first consider the effective DM-photon interaction from
\begin{align}\label{eq:lag}
\mathcal{L}\supset-\frac{1}{4} g_{a\gamma \gamma} aF_{\mu\nu} \tilde{F}^{\mu\nu}\,,
\end{align}
where $a$ denotes the pseudoscalar DM candidate with a dimensionful coupling $g_{a\gamma\gamma}$, and $\tilde{F}^{\mu\nu}$ denotes the dual of the photon field strength tensor $F^{\mu\nu}$.  This interaction can induce the $N\to2$ annihilation channel shown in Fig.~\ref{fig:NDM2}. For this process, we find that the amplitudes  satisfy the recursion relation
\begin{align}\label{recursion}
\mathcal{M}_{N+1}=\frac{g_{a\gamma\gamma}}{2}\left(1+\frac1N\right)^{N+2}\mathcal{M}_N\,,
\end{align}
which can be used to determine the large-$N$ amplitudes from small-$N$ ones. The derivation of this recursion relation is given in Appendix~\ref{app:recursion}. Parameterizing 
\begin{align}\label{eq:MN-def}
|\mathcal{M}_N|^2\equiv \kappa_Ng_{a\gamma\gamma}^{2N}m_{\rm dm}^4\,,
\end{align}
with $\kappa_N$ being dimensionless, we can infer from  \Eq{recursion} that  $\kappa_{N+1}=\kappa_N (1+1/N)^{2N+4}/4$, which leads to $\kappa_{N+1}\approx 1.85\,\kappa_N$ for large $N$. Given this, we first evaluate $|\mathcal{M}_6|^2$, with $\kappa_6=2657.21$  and then use $\kappa_N\approx 1.85^{N-6}\kappa_6$   to calculate  $|\mathcal{M}_N|^2$ at $N>6$.

We derive the bound on the DM-photon coupling by taking the largest $N$-body contribution from Eq.~\eqref{eq:DN_numerical}, yielding 
\begin{align}\label{gaaa_bound}
\log_{10}\left(\frac{g_{a\gamma\gamma}}{{\rm GeV}^{-1}}\right)<\left(1-\frac{5}{2N}\right)\log_{10}\left(\frac{m_{\rm dm}}{\rm eV}\right)-0.24+\frac{\log_{10}\left[(N-2)N!/N\right]}{2N}+\frac{4.21}{N}.
\end{align}
The strongest constraint typically appears at $N\gg 10$ for $m_{\rm dm}$ below 1 eV. As an example, for $m_{\rm dm}=10^{-14}~{\rm eV}$, the  bound is $g_{a\gamma\gamma}<7.9\times10^{-14}~{\rm GeV}^{-1}$, which is derived at $N=184$ and stronger than the conventional $2\to2$ contribution by around 9 orders of magnitude.

\begin{figure}[t]
	\centering
	\includegraphics[scale=0.4]{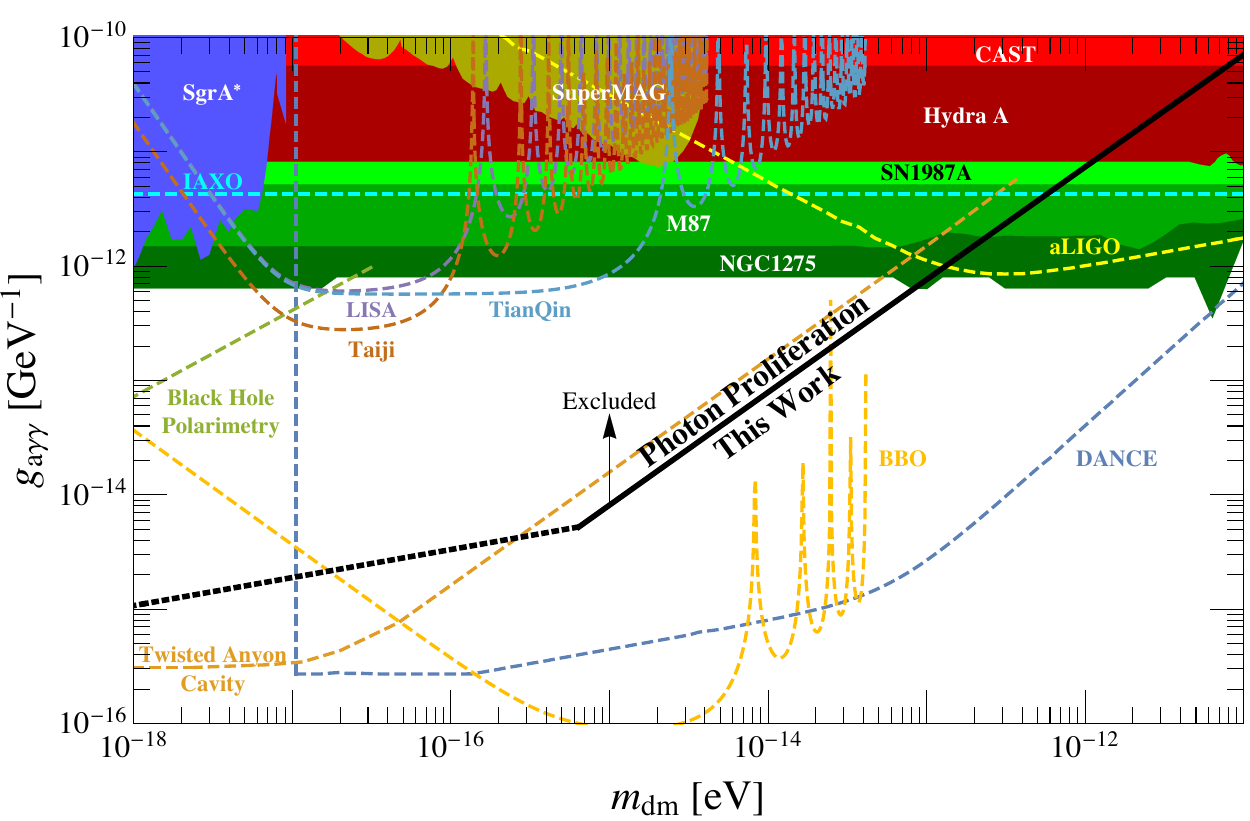}  
	\caption{\label{fig:gagg_bound} The upper bound of the  DM-photon coupling $g_{a\gamma\gamma}$  from the photon proliferation effect, which is induced from $N$-body DM annihilation via Fig.~\ref{fig:NDM2}. Current  bounds (shaded regions) and   detection limits from future   experiments (dashed lines)   are also shown for comparison. 
	}
\end{figure}

For each $m_{\rm dm}$, we derive the strongest $g_{a\gamma\gamma}$ constraint by varying $N$, and show it as the black line in Fig.~\ref{fig:gagg_bound}, the region above which is excluded by the $\Delta N_{\rm eff}$ constraint due to the photon proliferation effect. It should be noted that the constraint can vary in specific ultralight DM production scenarios. In the vacuum misalignment mechanism, for example, if the DM mass is sufficiently small, DM may not be nonrelativistic before neutrino decoupling, and Eq.~\eqref{deltan} must be modified accordingly. Concretely, the DM field starts to oscillate and is damped like nonrelativistic matter only after $H<m_{\rm dm}$. If this occurs after neutrino decoupling, the upper integration limit in Eq.~\eqref{deltan} would be adapted to $z_2[m_{\rm dm}/H(z_2)]^{1/2}$, where $H(z_2)=7.7\times 10^{-16}$~eV corresponds to  neutrino decoupling at $T=1.32$~MeV. In this case, the constraint becomes weaker than from Eq.~\eqref{gaaa_bound} when $m_{\rm dm}<H(z_2)$, as shown by the black dotted line.

Other ultralight DM production mechanisms beyond the traditional vacuum misalignment, such as those delaying the onset of ultralight DM oscillations through additional kinetic energy or gradient energy~\cite{Co:2017mop,Co:2019jts,Harigaya:2019qnl}, could also lead to DM becoming nonrelativistic long after neutrino decoupling, thereby weakening the $N_{\rm eff}$ constraints. However, in these scenarios, there is generally  significant kinetic energy contributing to  the evolution of the early Universe, which may cause early kination domination and hence  leads to significant deviations in the prediction of the standard BBN and CMB~\cite{Co:2021lkc}.  Besides, these active DM modes will contribute to warm or even hot DM components, suppressing the matter power spectrum through free streaming. In particular, ultralight DM becoming nonrelativistic below the keV temperatures would suffer from strong bounds from CMB spectral distortions~\cite{Chluba:2015hma} and Lyman-$\alpha$ forest~\cite{Harigaya:2025pox}. Consequently, such alternative DM production scenarios face additional stringent constraints even though the $N_{\rm eff}$ bound becomes irrelevant.



For $m_{\rm dm}\lesssim10^{-13}$ eV, the photon proliferation effect presents stronger constraints on the DM-photon coupling compared with the existing bounds shown in shaded regions, including  the observations from the Chandra mission (on NGC 1275~\cite{Reynolds:2019uqt}, M87~\cite{Marsh:2017yvc}, and Hydra A~\cite{Wouters:2013hua}), the SN1987A gamma-ray data~\cite{Hoof:2022xbe}, the CERN Axion Solar Telescope (CAST)~\cite{CAST:2017uph}, the SuperMAG~\cite{Arza:2021ekq,Friel:2024shg}, the Sgr A*~\cite{Yuan:2020xui}. The bound derived here also  rules out a large portion of detection windows in several future experiments (dashed lines), including  the Dark matter Axion search with riNg Cavity Experiment (DANCE)~\cite{Michimura:2019qxr}, the Twisted Anyon Cavity~\cite{Bourhill:2022alm}, the black hole polarimetry~\cite{Gan:2023swl}, the International AXion Observatory (IAXO)~\cite{Shilon:2012te}, and the gravitational wave detectors (aLIGO~\cite{Nagano:2019rbw} and space-based interferometers such as LISA, TianQin, Taiji, and BBO~\cite{Yao:2024hap}). Noticeably, the bound on $g_{a\gamma\gamma}$ derived here does not  assume  model-dependent relations between $g_{a\gamma\gamma}$ and $m_{\rm dm}$ (or the decay constant $f_a$ in some axion-like particle (ALP) models~\cite{Ringwald:2012hr}).

\subsection{The DM quartic self-coupling}

\begin{figure}[t]
	\centering
	\includegraphics[scale=0.55]{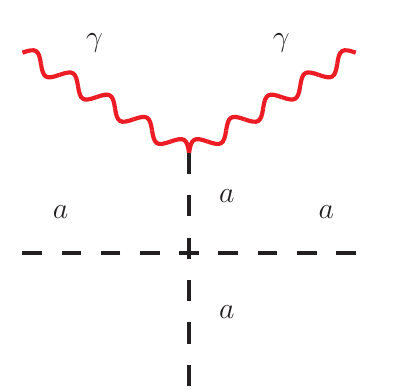} 
	\caption{\label{fig:NDM} The $3\to2$ annihilation process of DM to diphotons via the quartic self-coupling $\lambda$. The $N\geqslant 5$ channels can be induced by attaching more $\lambda$-vertices to the $a$-lines.
	}
\end{figure}

The discussions in the previous subsection assume that $N$-body DM annihilation to diphotons comes solely from the $a$-$\gamma$-$\gamma$ vertex. For pseudoscalar DM, the quartic self-coupling is usually also present either at tree-level or from quantum corrections, with the effective Lagrangian
\be\label{eq:self-coup}
\mL\supset-\frac{1}{4!}\lambda a^4\,.
\ee
The  basic $3\to2$ annihilation channel induced from Eq.~\eqref{eq:self-coup}  is  shown in Fig.~\ref{fig:NDM}, and the $N\to2$ annihilation processes with  $N\geqslant 5$ are generated by  attaching more $\lambda$-vertices to the DM lines, where  the squared amplitude  can be parameterized as 
\be\label{eq:M_Nprime}
|\mM'_N|^2\equiv\frac{\kappa'_N \lambda^{N-1} g_{a\gamma\gamma}^2}{m_{\rm dm}^{2N-6}}\, ,
\ee
with a dimensionless coefficient $\kappa'_N$.

Generally, for  self-interacting DM the $N\to2$ annihilation channel is induced by    Fig.~\ref{fig:NDM2}, Fig.~\ref{fig:NDM}, and the diagrams generated by attaching $\lambda$-vertices to DM lines in Fig.~\ref{fig:NDM2}. The resulting contribution to $|\Delta N_{\rm eff}|$ would contain Eq.~\eqref{eq:MN-def}, Eq.~\eqref{eq:M_Nprime}, and their interference terms. For a specific parameter choice of $(m_{\rm dm},g_{a\gamma\gamma},\lambda)$ and a given $N$, there might be a delicate cancellation effect from  the interference terms, which  leads to a very small $|\Delta N_{\rm eff}|$.  However, this is expected to be the tuning  case, and for a different  $N$  the cancellation would not appear.  Given that the modification to $N_{\rm eff}$ results from the sum of different $N\to 2 $ channels, we neglect the accidental cancellation and    derive the $\lambda$ bound  via the  $N\to2$ annihilation process induced from Fig.~\ref{fig:NDM}. 

Although a general formula for $\kappa'_N$ is difficult to obtain, we have used the packages \texttt{FeynArts}/\texttt{FeynCalc}~\cite{Hahn:2000kx, Shtabovenko:2023idz} to obtain $\kappa'_3=0.63$, $\kappa'_5=0.85$, and $\kappa'_7=2.49$, and find that $\kappa'_N$ slowly increases with $N$. Then for large $N$, we consider a conservative estimate of  $|\mM'_N|^2$ by taking $\kappa'_N=\kappa'_7$ for $N\geqslant7$.
Using the $|\Delta N_{\rm eff}|$ constraint from BBN and CMB, we obtain
\begin{multline}\label{eq:lamdaeff_bound}
\log_{10}\lambda<\frac{4N-7}{N-1}\log_{10}\left(\frac{m_{\rm dm}}{\rm eV}\right)-\frac{18.2N-27.8}{N-1}
\\
-\frac{2}{N-1}\log_{10}\left(\frac{g_{a\gamma\gamma}}{{\rm GeV}^{-1}}\right)+\frac{\log_{10}\left[(N-2)N!/N\right]}{N-1}. 
\end{multline}
The bound of $\lambda$ also depends on $g_{a\gamma\gamma}$, but the factor of  $2/(N-1)$ makes the dependence rather insensitive at large $N$. Given this, we substitute the bound of $g_{a\gamma\gamma}$ from Eq.~\eqref{gaaa_bound} into Eq.~\eqref{eq:lamdaeff_bound}. It turns out that the bound of $\lambda$ is not sensitive to the precise value of $N$, as the right-hand side changes slowly with $N$. For $10^{-15}~{\rm eV}\lesssim m_{\rm dm}\lesssim10^{-6}~{\rm eV}$, the upper limit of $\lambda$ varies  from $\mO(10^{-76})$ to $\mO(10^{-40})$, which is    much stronger than  from CMB anisotropies~\cite{Cembranos:2018ulm}. It is worth mentioning that  the strongest  constraint  on $\lambda$ is derived    at $N=\mathcal{O}(100)$, and  the   bound derived at   $N=7$ is   weaker by 1 (5) order(s) of magnitude  at $m_{\rm dm}=10^{-6}~(10^{-15})$~eV.  

\begin{figure}[t]
\centering
\includegraphics[scale=0.4]{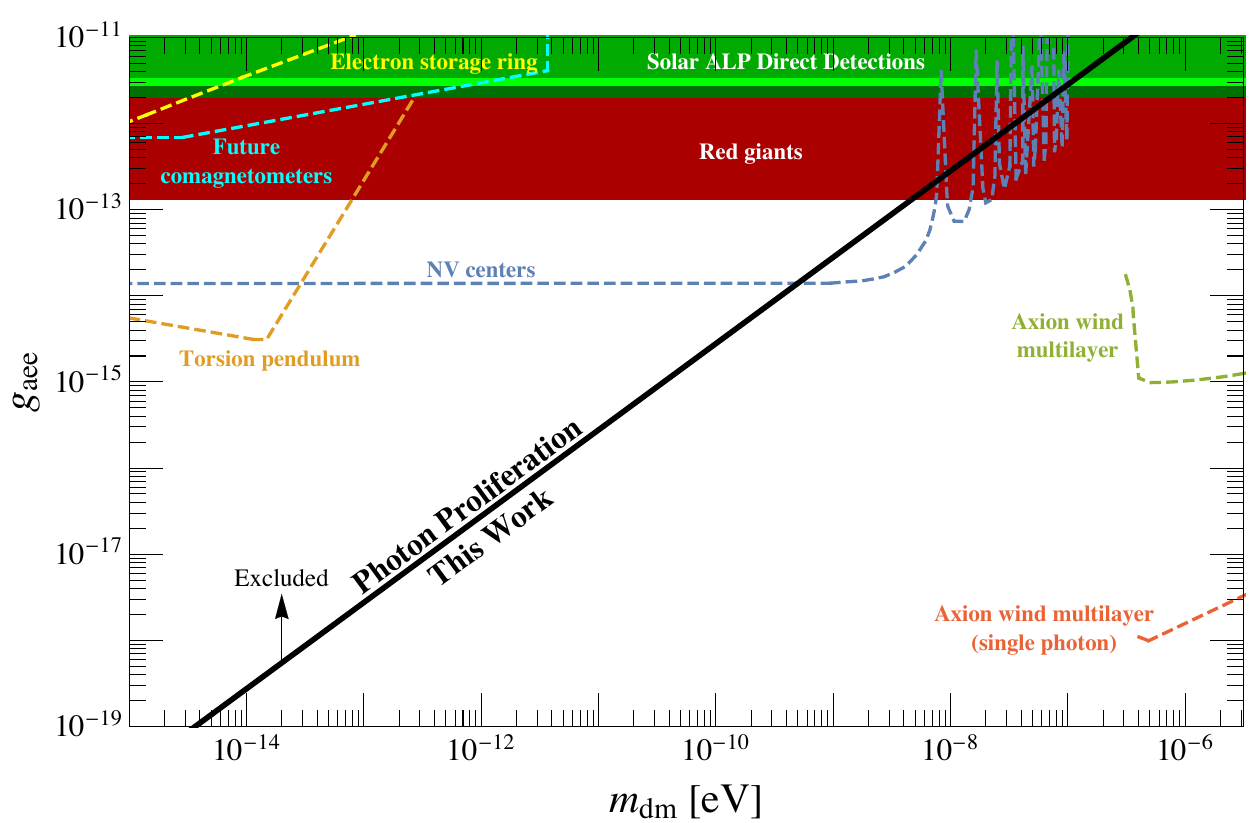}  
\caption{The upper  bound on the DM-electron coupling $g_{aee}$, where the photon proliferation effect arises from the  $g_{aee}$-inducced quartic DM coupling. Current  bounds (shaded regions)  and future detection sensitivities (dashed lines)   are   shown for comparison.}\label{fig:lambda_eff}
\end{figure}

In the presence of interaction between pseudoscalar DM and electrons, the generic coupling can be described by shift-symmetric interaction $-m_e e^{2ia/f}\bar e_R e_L+\rm h.c.$, where $m_e$ is the electron mass and $f$ corresponds to  some symmetry breaking scale. Up to $\mathcal{O}(a^2/f^2)$, the low-energy effective Lagrangian reads
\begin{align}\label{eq:DM-e_lag}
\mathcal{L}\supset-im_e \frac{a}{f}  \bar e \gamma_5 e+m_e \frac{a^2}{f^2} \bar e e\,.
\end{align}
Usually, only the first term is considered in constraints of light particles coupling to  electrons. However, the second term is necessary to ensure the mass stability of the ultralight DM particle  under self-energy corrections~\cite{Bauer:2023czj}. At one-loop level, \Eq{eq:DM-e_lag} induces an effective quartic coupling for DM self-interaction,
\begin{align}
\lambda_{\rm eff}=\frac{24}{\pi^2}g_{aee}^4\ln\left(\frac{m_e^2}{m_{\rm dm}^2}\right)\,,
\end{align}
where $g_{aee}\equiv m_e/f$, and the $\overline{\text{MS}}$ renormalization scheme is used with the running scale fixed at $m_{\rm dm}$.

Following \Eq{eq:lamdaeff_bound}, we can obtain an upper bound on $g_{aee}$ via $\lambda_{\rm eff}$, as shown in the black line of Fig.~\ref{fig:lambda_eff}. For $10^{-15}~{\rm eV}\lesssim m_{\rm dm}\lesssim10^{-6}~{\rm eV}$, the limit of $g_{aee}$ varies from $\mO(10^{-20})$ to $\mO(10^{-11})$. Existing bounds from solar ALP search on the DM direct detection experiments (PandaX~\cite{PandaX-II:2020udv,PandaX:2024cic}, LUX~\cite{LUX:2017glr}, XENONnT~\cite{XENON:2022ltv}) and from the red giant observations~\cite{Giannotti:2017hny} are shown in shaded regions. It is seen that the photon proliferation  effect presents stronger bounds on $g_{aee}$ than the current limits  at $m_{\rm dm}\lesssim\mathcal{O}(10^{-8})$~eV. We also show the projected detection sensitivities on $g_{aee}$ from the electron storage ring~\cite{Brandenstein:2022eif}, the future comagnetometer~\cite{Bloch:2019lcy}, the Nitrogen-Vacancy centers in diamonds~\cite{Chigusa:2023hms}, the spin precession experiment via torsion pendulum~\cite{Graham:2017ivz}, and the axion wind multilayer~\cite{Berlin:2023ubt}. As can be seen from Fig.~\ref{fig:lambda_eff}, a large portion of the parameter space targeted by these future experiments is already excluded.

\section{Conclusion}

We have shown in this paper that multi-body nonrelativistic DM annihilation  can  dominate over the conventional two-body process in the early Universe, which can be realized in a wide range of nonthermal DM scenarios. As a typical example,  we demonstrate a novel photon proliferation effect from $N$-body pseudoscalar DM annihilation to diphotons, which is a  general phenomenon for ultralight DM that couples to photons and has become nonrelativistic at MeV temperatures. We also illustrated that for DM mass approaching the ultralight end, the photon proliferation effect  will present the leading constraints on the DM-photon, DM self-interaction, and DM-electron couplings. The discussion can be extended to other interactions such as  the DM-neutrino interaction, and other  DM candidates like the dark photon. Our results highlight the importance of $N \to 2$ annihilation in the early Universe when physical impacts are boosted, and open  new avenues to probe hidden particles once abundant in the early Universe but disappeared in the present day.

As final remarks, we would like to summarize the working regime for the constraints we have derived. For both $g_{a\gamma\gamma}$ coupling and self-interacting coupling $\lambda$, the DM relic density we used for the ultralight DM particle is from the present-day value dated back to earlier times, focusing on temperatures below $1$~MeV. We assumed that the ultralight DM particle has already become nonrelativistic at this epoch. This is the general situation for axion or axion-like DM candidates following the standard vacuum misalignment mechanism with $m_{\rm dm}
\gtrsim10^{-15}$~eV. For the vacuum misalignment with $m_{\rm dm}
\lesssim10^{-15}$~eV or other mechanisms that delay the onset of DM oscillations until the cosmic temperature drops far below 1~MeV, our $N_{\rm eff}$ bound weakens due to the change of upper integration limit in \Eq{deltan}. However, in such late-production scenarios, one must ensure DM becomes cold enough  before structure formation and should not create large  photon spectral distortions, which deserves careful case-by-case considerations.

\acknowledgments

We would like to thank Amin Aboubrahim, Jinhui Guo, Jia Liu, Yong Tang, Jingqiang Ye, and Bingrong Yu for useful discussions on ultralight DM,  Vladyslav Shtabovenko for helpful comments on {\tt FeynArts}/{\tt FeynCalc}, and Jens Chluba for discussions on CMB spectral distortions from large photon number injection.  We appreciate the helpful data repositories of ultralight DM constraints managed by Ciaran A.~J.~O'Hare~\cite{AxionLimits}. K.-P.~Xie also thanks the hospitality of Osaka University where part of this work was performed. S.-P.~Li is  supported by JSPS Grant-in-Aid for JSPS Research Fellows No. 24KF0060. K.-P.~Xie is supported by the Fundamental Research Funds for the Central Universities.

\appendix
\section{Squared amplitudes from photon-mediated annihilation}\label{app:recursion}

Here we will  demonstrate the recursion relation between $|\mathcal{M}_{N+1}|^2$ and $|\mathcal{M}_{N}|^2$ from the fence diagrams shown in Fig.~1 of the main text. Defining the $N\to 2$ squared amplitude as
\begin{align}\label{eq:MN_def}
|\mathcal{M}_N|^2\equiv\kappa_Ng_{a\gamma\gamma}^{2N} m_{\rm dm}^4\,,
\end{align} 
we are to find out the  relation between the dimensionless coefficients $\kappa_{N+1}$ and  $\kappa_{N}$,  and in particular show that $\kappa_{N+1}>\kappa_{N}$.

\begin{figure}[t]
	\centering
	\includegraphics[scale=0.45]{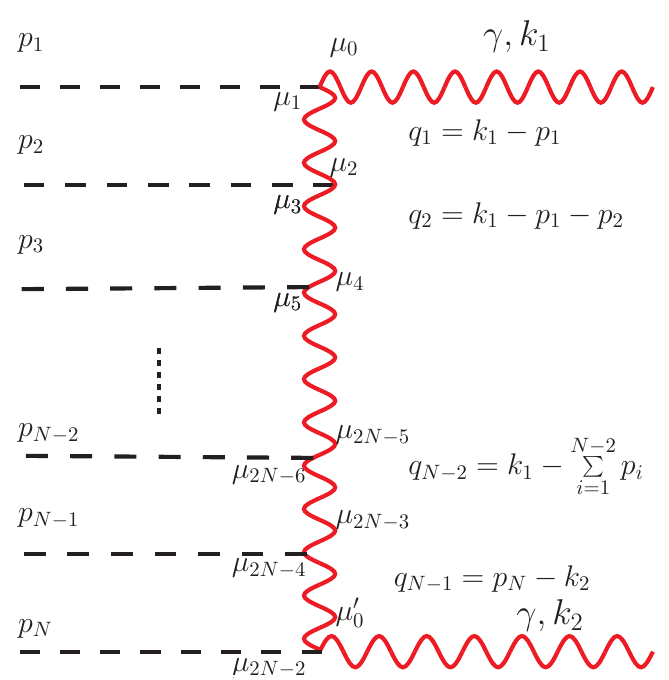} 
	\caption{\label{fig:NDM_full}   $N$-body DM annihilation to diphotons via the effective DM-photon coupling, where $\mu_i, \mu'_0$ are used to denote the Lorentz indices in the DM-photon-photon vertices. 
	}
\end{figure}

From Fig.~\ref{fig:NDM_full}, we can write down the general amplitude in the limit of nonrelativistic DM annihilation:
\begin{multline}\label{eq:amp_gen_2}
\mathcal{M}_N=g_{a\gamma\gamma}^N N!\left(\prod_{ i=1}^{N-1}\frac{\eta_{\mu_{2i-1}\mu_{2i}}}{(i^2-iN)m_{\rm dm}^2}\right) \left(\prod_{j=2}^{N-1}\epsilon^{q_{j-1}q_{j} \mu_{2j-2}\mu_{2j-1}}\right)\\
\times\epsilon^{k_1 q_1 \mu_0\mu_1}\epsilon^{q_{N-1}k_2\mu_{2N-2}\mu'_0}\epsilon_{\mu_0}^*(k_1)\epsilon_{\mu'_0}^*(k_2)\,,
\end{multline}
where we have dropped a potential minus sign in the amplitude, and contraction  between the Levi-Civita tensor and momenta is defined as
\begin{equation}\begin{split}
\epsilon^{k_1 q_1 \mu_0\mu_1}\equiv&~\epsilon^{\alpha_0 \alpha_1 \mu_0 \mu_1}k_{1, \alpha_0}q_{1,\alpha_1}\,,
\\
\epsilon^{q_{N-1}k_2\mu_{2N-2}\mu'_0}\equiv&~\epsilon^{\alpha_{2N-2}\alpha'_0 \mu_{2N-2} \mu'_0}q_{N-1,\alpha_{2N-2}}k_{2, \alpha'_0} \,,
\\
\epsilon^{q_{i-1}q_{i} \mu_{2i-2}\mu_{2i-1}}\equiv&~\epsilon^{\alpha_{2i-2}\alpha_{2i-1}\mu_{2i-2} \mu_{2i-1}}q_{i-1, \alpha_{2i-2}} q_{i,\alpha_{2i-1}}\,,
\end{split}\end{equation} 
with $\epsilon^{0123}=1$. The  kinetics for $N\to 2$ annihilation in the center-of-mass frame gives
\be\label{eq:kinetics}\begin{split}
k_1=&~\frac{N}{2}m_{\rm dm}\left(1, \vec e_k\right)\,,\quad k_2=\frac{N}{2}m_{\rm dm}\left(1, -\vec e_k\right)\,,\\
q_i=&~m_{\rm dm}\left(\frac{N-2i}{2}, \frac{N}{2} \vec e_k\right),\quad\text{for $i\leqslant N-1$}\,,
\end{split}\ee
where $\vec e_k$ denotes the unit vector of  the outgoing photon momentum $\vec k_1$, and we take $\vec e_k\equiv (0,0,1)$ without loss of generality. Furthermore, we have $k_1^2=k_2^2=0$ and $q_i^2=(i^2-i N)m_{\rm dm}^2$.

For $N+1\to 2$ annihilation, it is straightforward to obtain the kinetics by replacing  $N$ with $N+1$ in Eq.~\eqref{eq:kinetics}. Since we intend to derive the recursion relation, now we express the kinetics  (primed) from $N+1\to 2$ annihilation in terms of  those (unprimed) from $N\to 2$ annihilation:
\begin{align}
	k'_1=r_1+k_1\,, \quad k'_2=r_2+k_2\,,\quad q'_i=r_1+q_i \, (i\leqslant N-1)\,,\quad  q'_N=-r_2+q_{N-1}\,,
\end{align}
with  
\be\label{eq:kinetics_new}\begin{split}
r_1=&~\frac{m_{\rm dm}}{2}\left(1,\vec e_k\right),\quad r_2=\frac{m_{\rm dm}}{2}\left(1, -\vec e_k\right),\\
q^{\prime 2}_i=&~q_i^2\left(1-\frac{i}{i^2-i N}\right), \quad q^{\prime 2}_N=-N m_{\rm dm}^2\,.
\end{split}\ee
Using the above  relations and Eq.~\eqref{eq:amp_gen_2}, we can write down the general amplitude for $N+1\to 2$ annihilation, 
\begin{align}\label{eq:amp_gen_N+1}
	\mathcal{M}_{N+1}&= \left(\frac{g_{a\gamma\gamma}(N+1)}{N m_{\rm dm}^2}\right) \prod_{k=1}^{N-1}\left(1-\frac{k}{k^2-k N}\right)^{-1} 
	\\
	&\times  \eta_{\mu_{2N-1}\mu_{2N}} \epsilon^{(r_1+k_1) (r_1+q_1) \mu_0\mu_1} \epsilon^{(q_{N-1}-r_2)(r_2+k_2)\mu_{2N}\mu'_0} \epsilon^{(r_1+q_{N-1}) (q_{N-1}-r_2) \mu_{2N-2}\mu_{2N-1}}
	\nonumber\\
	& \times g_{a\gamma \gamma}^{N}N! \left(\prod_{ i=1}^{N-1}\frac{\eta_{\mu_{2i-1}\mu_{2i}}}{(i^2-iN)m_{\rm dm}^2}\right) \left(\prod_{j=2}^{N-1}\epsilon^{(r_1+q_{j-1}) (r_1+q_j) \mu_{2j-2}\mu_{2j-1}}\right)  \epsilon_{\mu_0}^*(k_1)\epsilon_{\mu'_0}^*(k_2)\,,\nn
	\end{align}
where the first line is a new term from $N+1\to 2$ annihilation, while the second and third lines contain both the  $N\to 2$ terms and new contributions   from  $r_1, r_2$.  The first line gives an overall factor 
\begin{align}
\left(\frac{g_{a\gamma\gamma}(N+1)}{N m_{\rm dm}^2}\right) \prod_{k=1}^{N-1}\left(1-\frac{k}{k^2-k N}\right)^{-1} =g_{a\gamma\gamma}\frac{N+1}{N^2 m_{\rm dm}^2}\,.
\end{align}
The first contraction in the second line reads
\begin{equation}\begin{split}\label{eq:eps_mu0mu1}
\epsilon^{(r_1+k_1) (r_1+q_1) \mu_0\mu_1}=&~\epsilon^{\alpha_0 \alpha_1 \mu_0\mu_1} \left(r_{1,\alpha_0}q_{1,\alpha_1}+k_{1,\alpha_0}r_{1,\alpha_1}+k_{1,\alpha_0}q_{1,\alpha_1}\right)\\
=&~\epsilon^{\alpha_0 \alpha_1 \mu_0\mu_1}\left(k_{1,\alpha_0}q_{1,\alpha_1}-r_{1,\alpha_0}p_{1,\alpha_1}\right)
=\left(1+\frac{1}{N}\right)\epsilon^{k_1 q_1 \mu_0\mu_1}\,,
\end{split}\end{equation}
where $\epsilon^{\alpha_0 \alpha_1 \mu_0\mu_1}=-\epsilon^{ \alpha_1 \alpha_0 \mu_0\mu_1}$ and $q_1=k_1-p_1$ were used in the second equation, and the last result was derived by  using the kinetics given in Eqs.~\eqref{eq:kinetics}--\eqref{eq:kinetics_new}. This result indicates that the first term in the second equation of Eq.~\eqref{eq:eps_mu0mu1}, which comes from    $N\to 2$ annihilation,  is at $\mathcal{O}(Nm_{\rm dm}^2)$, while the second term, which  is a new contribution in $N+1\to 2$ annihilation, is   at $\mathcal{O}(m_{\rm dm}^2)$. 
The second contraction in the second line of Eq.~\eqref{eq:amp_gen_N+1} also exhibits the similar feature, where
\bea\label{eq:eps_m2N1m0p}
\epsilon^{(q_{N-1}-r_2)(r_2+k_2)\mu_{2N}\mu'_0}&=&\epsilon^{\alpha_{2N}\alpha'_0\mu_{2N}\mu'_0}\left(q_{N-1,\alpha_{2N}}k_{2,\alpha'_0}-r_{2,\alpha_{2N}}k_{2,\alpha'_0}+q_{N-1,\alpha_{2N}}r_{2,\alpha'_0}\right)
\nn\\
&=&\left(1+\frac{1}{N}\right)\epsilon^{q_{N-1} k_2 \mu_{2N}\mu'_0}\,.
\eea
The first term  in the first line of Eq.~\eqref{eq:eps_m2N1m0p}  would correspond to $N\to 2$ annihilation up to a different index from $\mu_{2N}$. By using the property of the Levi-Civita tensor,  the second and third terms can be verified to be:
\begin{align} \label{eq:eps_m21N1m0p}
\epsilon^{\alpha_{2N}\alpha'_0\mu_{2N}\mu'_0}\left(-r_{2,\alpha_{2N}}k_{2,\alpha'_0}+q_{N-1,\alpha_{2N}}r_{2,\alpha'_0}\right)&=-\frac{m_{\rm dm}^2}{2}	\epsilon^{03\mu_{2N}\mu'_0}\,,
\end{align}
where the first two indices in the Levi-Civita tensor  result from the fact that only the time-space and space-time components survive from the 4-momentum product and the index 3 appears since we have taken the spatial momentum in the $z$-direction.  By comparing Eq.~\eqref{eq:eps_m2N1m0p} with the corresponding term in $N\to 2$ annihilation,  we find that  in addition to   a different index from $\mu_{2N}$,  an overall factor arises  when going from $N\to 2$ to $N+1\to 2$ annihilation. 

By similar arguments,  we find that the  second product in the third line of Eq.~\eqref{eq:amp_gen_N+1} reads
\bea
&&\prod_{j=2}^{N-1}\epsilon^{(r_1+q_{j-1}) (r_1+q_j) \mu_{2j-2}\mu_{2j-1}}
\nn \\
&=&\prod_{j=2}^{N-1}\epsilon^{\alpha_{2j-2}\alpha_{2j-1} \mu_{2j-2}\mu_{2j-1}} \left(q_{j-1,\alpha_{2j-2}} q_{j, \alpha_{2j-1}}-r_{1,\alpha_{2j-2}} p_{j, \alpha_{2j-1}}\right)\nn\\
&=&\left(1+\frac{1}{N}\right)^{N-2}\prod_{j=2}^{N-1}\epsilon^{q_{j-1}q_j\mu_{2j-2}\mu_{2j-1}}\,,
\eea
where $q_j-q_{j-1}=-p_j$ was used in the first equation. Therefore, an overall factor arises from this  product when going from $N\to 2$ to $N+1\to2$ annihilation.

Finally, the third contraction in the second line of Eq.~\eqref{eq:amp_gen_N+1} is a new term from $N+1\to 2$ annihilation. Using the kinetics, we find that 
\begin{equation}\begin{split}
&~\epsilon^{(r_1+q_{N-1}) (q_{N-1}-r_2) \mu_{2N-2}\mu_{2N-1}}\\
=&~\epsilon^{\alpha_{2N-2}\alpha_{2N-1} \mu_{2N-2}\mu_{2N-1}}\left[\left(r_1+r_2\right)_{\alpha_{2N-2}} q_{N-1, \alpha_{2N-1}}-r_{1,\alpha_{2N-2}}r_{2,\alpha_{2N-1}}\right]\\
=&~\frac{(N+1)m_{\rm dm}^2}{2}\epsilon^{03\mu_{2N-2}\mu_{2N-1}}\,,
\end{split}\end{equation}
where the reason for the appearance of the 03 indices is the same as that in Eq.~\eqref{eq:eps_m21N1m0p}.  

Assembling these terms, we can rewrite the $N+1\to 2$ amplitude as
\begin{align}\label{eq:amp_gen_N+1_2}
	\mathcal{M}_{N+1}&= \left(\frac{g_{a\gamma\gamma}}{2}\right) \left(\frac{N+1}{N}\right)^{2+N}\eta_{\mu_{2N-1}\mu_{2N}}\epsilon^{03 \mu_{2N-2}\mu_{2N-1}}\epsilon^{k_1 q_1 \mu_0\mu_1}\epsilon^{q_{N-1}k_2\mu_{2N}\mu'_0}
	\nonumber\\[0.2cm]
	& \times g_{a\gamma \gamma}^{N}N! \left(\prod_{ i=1}^{N-1}\frac{\eta_{\mu_{2i-1}\mu_{2i}}}{(i^2-iN)m_{\rm dm}^2}\right) \left(\prod_{j=2}^{N-1}\epsilon^{q_{j-1} q_j \mu_{2j-2}\mu_{2j-1}}\right)  \epsilon_{\mu_0}^*(k_1)\epsilon_{\mu'_0}^*(k_2)\,.
\end{align}
Comparing to Eq.~\eqref{eq:amp_gen_2}, we find that an overall factor arises in the $N+1\to 2$ annihilation $(1+1/N)^{2+N}/2$ in addition to the expected DM-photon coupling  $g_{a\gamma\gamma}$.  The remaining difference can be parameterized as
\begin{align}
N\to 2 &: \mathcal{M}_{N, \mu_{2N-2}}\epsilon_{\mu'_0}^*(k_2)\epsilon^{q_{N-1}k_2\mu_{2N-2}\mu'_0}\,,
\\[0.2cm] 
N+1\to 2 &: \mathcal{M}_{N+1, \mu_{2N-2}}\epsilon_{\mu'_0}^*(k_2)\eta_{\mu_{2N-1}\mu_{2N}}\epsilon^{03 \mu_{2N-2}\mu_{2N-1}}\epsilon^{q_{N-1}k_2\mu_{2N}\mu'_0}\,.
\end{align}
To see the above difference, we  recall that only the  03 and 30 components in the first two indices of  the Levi-Civita tensors ($\epsilon^{q_{N-1}k_2\mu_{2N-2}\mu'_0}$ and $\epsilon^{q_{N-1}k_2\mu_{2N}\mu'_0}$) survive. Then we  perform the  contraction of index $\mu'_0$, with  the  photon polarization vector $\epsilon^*_{\mu'_0}=(0,0,1,0)$, which selects $\mathcal{M}_{N,1}$ for $N\to2$ annihilation and $\mathcal{M}_{N+1,2}$ for $N+1\to2$ annihilation. These two results, when further repeatedly contracting with other indices  via the Minkowski metric tensor $\eta_{\mu_{i}\mu_{j}}$, will be identical up to a potential difference of a total  minus sign. Therefore, we arrive at the final recursion relation
\begin{align}
	\mathcal{M}_{N+1}=\left(\frac{g_{a\gamma\gamma}}{2}\right) \left(\frac{N+1}{N}\right)^{2+N}\mathcal{M}_N\,,
\end{align}
and consequently the relation
\begin{align}
\kappa_{N+1}=\frac{1}{4}\left(\frac{N+1}{N}\right)^{2N+4}\kappa_N\,.
\end{align}
We can easily check this relation by using the results obtained from the  \texttt{FeynArts}/\texttt{FeynCalc} packages~\cite{Hahn:2000kx, Shtabovenko:2023idz}. The squared amplitude of the $1\to 2$ and  $2\to2$ processes are $|\mathcal{M}_{1}|^2=g_{a\gamma\gamma}^2m_{\rm dm}^4/2$ and $|\mathcal{M}_{2}|^2=8 g_{a\gamma\gamma}^4m_{\rm dm}^4$, and the ones up to $N=6$ read
\bea
|\mathcal{M}_{3}|^2=\frac{6561}{128} g_{a\gamma\gamma}^6 m_{\rm dm}^4\,, ~\,
&|\mathcal{M}_{4}|^2=\frac{2048}{9} g_{a\gamma\gamma}^8m_{\rm dm}^4\,, \nn\\[0.2cm]
|\mathcal{M}_{5}|^2=\frac{244140625}{294912} g_{a\gamma\gamma}^{10}m_{\rm dm}^4\,,~\,
&|\mathcal{M}_{6}|^2= \frac{531441}{200} g_{a\gamma\gamma}^{12}m_{\rm dm}^4\,,
\eea
where $\kappa_N$ can be extracted via Eq.~\eqref{eq:MN_def}. We have also checked that the above results can be consistently derived by the general amplitude~\eqref{eq:amp_gen_2}.

\bibliographystyle{JHEP-2-2.bst}
\bibliography{Refs}

\end{document}